\documentstyle[11pt,paspconf,epsf]{article}

\begin{document}

\title{Effects of Grain Morphology and Impurities on the Infrared Spectra 
of Silicon Carbide Particles}
\author{H. Mutschke, Th. Henning and D. Cl\'ement}
\affil{Astrophysical Institute and University Observatory, Friedrich Schiller 
University, Schillerg\"a\ss{}chen 3, D-07745 Jena, Germany}
\author{A.C. Andersen}
\affil{Niels Bohr Institute, Astronomical Observatory,Copenhagen University,
     Juliane Maries Vej 30, DK-2100 Copenhagen, Denmark}

\begin{abstract}
In this paper we demonstrate that distinguishing between the polytypes 
of silicon carbide by means of infrared features in small-grain spectra 
is impossible. Therefore, the infrared spectra of carbon stars, 
unfortunately, do not provide a means for drawing comparisons between 
the crystal structures of grains condensed in these environments and 
found in meteorites. 

This is proven first by comparing theoretical band profiles calculated 
for ellipsoidal particles, which show clearly a strong dependence 
on the axis ratio of the ellipsoids but negligible differences for the 
two most common polytypes. Second, spectra measured on submicron particle 
samples in the laboratory do not show any obvious correlation of band 
position or shape to the polytype. However, we demonstrate by measurements 
on SiC whiskers that grain shape is able to determine the spectrum 
completely. A further strong systematic influence on the band profile 
can be exerted by plasmon-phonon coupling due to conductivity of the 
SiC material. The latter fact probably is responsible for the confusion 
in the astronomical literature about spectral properties of SiC grains. 
We show that, although the conductivity seems to be a common property 
of many SiC laboratory samples, it is, however, independent of the 
polytype. \\[-1cm]
\end{abstract}


\section{Introduction}

Thermal emission of solid silicon carbide (SiC) particles has been observed 
in the spectra of circumstellar envelopes around carbon stars 
(e.g. Little-Marenin 1986, Speck et al.~1997). The 
identification is based on a broad infrared feature in the 
10-13~$\mu$m wavelength range, which is ascribed to the fundamental phonon 
transition of SiC. Attempts have been made to derive the crystal type 
of the circumstellar SiC grains from a comparison of the observed 
band profiles to those measured on laboratory analogs 
(Speck et al.~1997, 1999). 
Although knowledge about the crystal types (polytypes) of cosmic SiC 
grains would be very helpful, four points must be objected to such a 
comparison and the conclusions drawn from it:
\begin{enumerate}
\item The number of laboratory samples on which the comparison 
relies is very small and the samples have not been characterized 
sufficiently. In Sect.\,3 we will show laboratory spectra of 
another set of SiC samples, which fully contradict the conclusions 
drawn by Speck et al.\,(1999).
\item The phonon frequencies of the several SiC polytypes are 
experimentally and theoretically known to be very similar 
(Mutschke et al.\,1999). As we will show in Sect.\,2 by 
calculating SiC particle spectra, therefore very similar band 
profiles are expected for the different polytypes.
\item As we have reviewed in another paper in this volume 
(Henning \& Mutschke), absorption and emission by small grains near 
such strong transitions as the single-phonon band in SiC crystals 
is entirely determined by surface modes. Therefore, the grain morphology 
(shape, size and agglomeration) has a huge influence on the spectrum. 
This will be demonstrated by the calculations in Sect.\,2 as well as 
by measurements (Sect.\,4).
\item The $\beta$-SiC laboratory spectra used in the comparisons 
by Speck et al.\,(1997, 1999) are probably influenced by impurity-induced 
conductivity, which via plasmon-phonon coupling determines the 
appearance of the phonon band profile. Unfortunately, such impurity 
effects are not restricted to one or the other polytype. 
In Sect.\,5 we will discuss this effect by means of a number of new 
laboratory spectra. 
\end{enumerate}

\section{Theoretical Predictions for SiC Particle Spectra}

SiC is one of the textbook examples for which the dielectric function 
in the phonon band can be described very exactly by a single 
Lorentzian oscillator. The oscillator parameters differ slightly from 
one polytype to another and for the anisotropic polytypes also 
for the polarizations in the directions of the different principal axes 
of the crystals. Mutschke et al.\,(1999) derived average values of the 
oscillator parameters for the most important polytypes based on an extensive 
literature study. 

We have used these parameters to calculate absorption spectra of particles 
in the Rayleigh limit (nanometer-sized particles, for an exact definition 
see Henning \& Mutschke, this volume) for ellipsoidally shaped grains 
composed of the two most common polytypes. We denote the cubic 3C polytype 
by $\beta$-SiC and the hexagonal 6H type by $\alpha$-SiC (see Mutschke et 
al.\,1999). The calculated spectra clearly show (Fig.\,1) that (i) there 
is no significant difference caused by the crystal type and (ii) the 
spectrum is very different for different grain shapes. 
In the case of uniform ellipsoidal (or spherical) shapes the extinction 
is dominated by resonances appearing between the transverse 
(about 12.6~$\mu$m) and longitudinal (about 10.3~$\mu$m) optical phonon 
frequencies. These resonances are related to the principal axes of the 
ellipsoids and are called ``surface modes'' (compare Bohren \& Huffman 
1983, chap.12). In the limiting case of infinite elongation of a particle 
axis (e.g. for extreme needle- or plate-like shapes) an additional very 
strong resonance occurs at the transverse optical frequency corresponding 
to the volume mode of the material (see Fig.\,2). 

Unfortunately, theory at the moment is not able to calculate the extinction 
for other more realistic grain shapes or distributions of shapes in strong 
absorption bands. It is quite clear that such a realistic sample will 
show a more continuous broad band covering more or less the whole 
wavelength range between the transverse and longitudinal optical phonon 
frequencies. A possibility for approximating such spectra is the assumption 
of a distribution of ellipsoidal shapes. A very extreme one with equal 
probability of all shapes (CDE, Bohren \& Huffman 1983) is shown 
in Figs.\,1 and 2. Similarly to the other calculated spectra it does 
not reveal significant differences between the polytypes, which might be 
observable in space. Although any measured spectrum could be, in principle, 
translated to such a distribution of ellipsoidal shapes (Papoular et al. 
1998), this certainly is not an exact approach especially for irregular, 
sharp-edged grains and should be handled with care. 

\begin{figure}
\plotfiddle{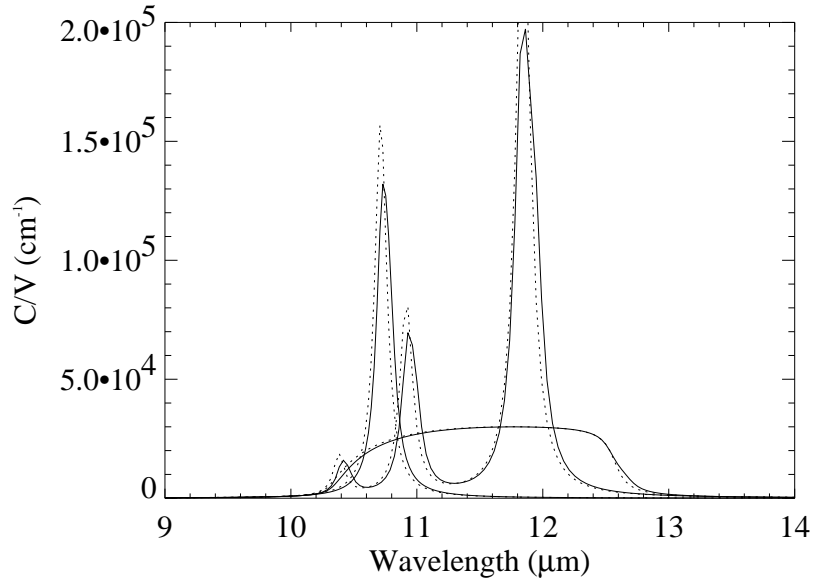}{3in}{0}{70}{70}{-150}{0}
\caption{Calculated extinction cross sections of spherical and 
ellipsoidal $\alpha$- (solid lines) and $\beta$-SiC (dotted lines) 
submicron particles. 
The single band with maximum at 10.7~$\mu$m is produced by 
spherical grains, the triple structure results from ellipsoidal shapes 
with an axis-ratio of 1:3:9, and the broad band has been calculated 
for a continuous distribution of ellipsoidal grains (CDE, 
after Bohren \& Huffman 1983)} \label{fig-1}
\end{figure}

The result for infinite plate-like particles (Fig.~2) has to be equivalent 
to usual thin-film spectra. The latter indeed are known to show the 
pure volume mode (12.6~$\mu$m) in transmission measurements normal 
to the interfaces, and an additional resonance at the position of 
the longitudinal optical frequency (10.3~$\mu$m) for a measurement 
at grazing incidence and polarization normal to the interfaces (the 
Berreman effect, Sciacca et al.\,1995). 
The spectra of the ``thin films'' produced by pressing SiC powder in 
a diamond anvil cell by Speck et al.\,(1999) do not show this typical 
thin-film characteristics, but a broad band quite similar to the CDE 
calculation and our measured particle spectra (see next sections). 
This indicates that forming a continuous thin film in the anvil cell 
fails in the case of the SiC powders, which together with 
a basic misunderstanding lead to questions about the influence of the 
particle-surrounding medium. The comparability of the anvil cell 
method with single-crystal transmission spectra, which is undoubted 
if real thin films are formed and only volume modes are measured, 
should not be intermixed with small-particle extinction measurements 
where surface modes dominate the spectrum.

\begin{figure}
\plotfiddle{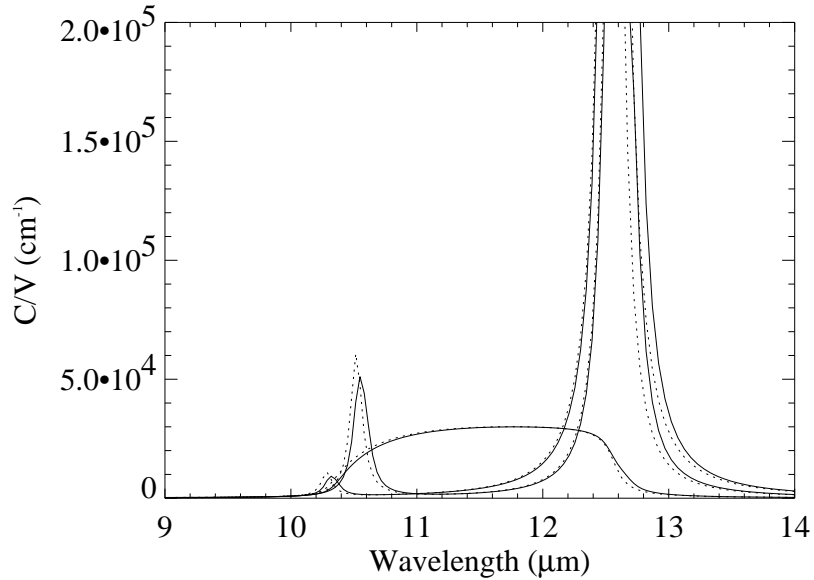}{3in}{0}{70}{70}{-150}{0}
\caption{Same as Fig.\,1 for needle- (resonance at 10.5~$\mu$m) 
and plate-like particles (resonance at 10.3~$\mu$m). The CDE (broad band) 
is again shown for comparison. $\alpha$-SiC .. solid lines, $\beta$-SiC 
.. dotted lines.} \label{fig-2}
\end{figure}

\section{Measurements on Polytypes}

\begin{figure}
\plotfiddle{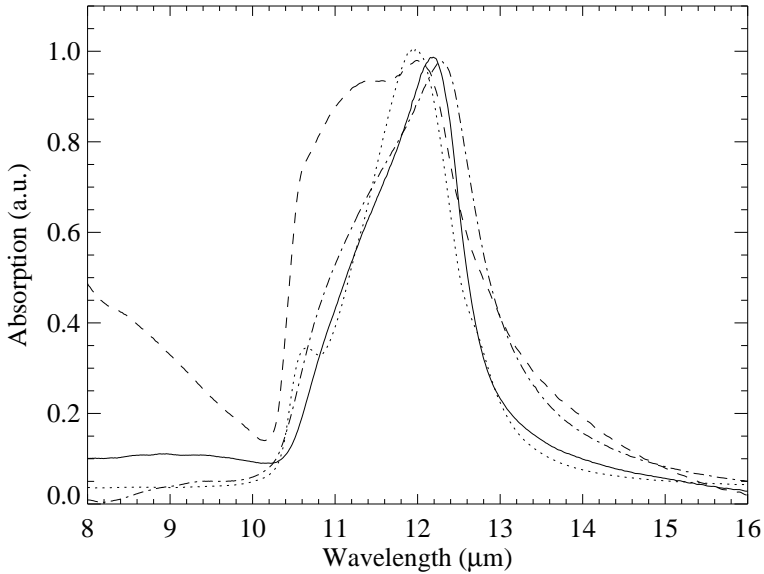}{3in}{0}{70}{70}{-150}{0}
\caption{Measured phonon band profiles for two $\alpha$- (dotted 
and dashed lines) and two $\beta$-SiC (solid and dash-dotted lines) samples.} 
\label{fig-3}
\end{figure}

As we have seen in the preceding section, from a theoretical point of 
view, spectra of SiC particles of different polytype should not differ 
very much - provided that the morphologies are the same. Nevertheless, 
spectra of the cubic ($\beta-$) and hexagonal/rhombohedral ($\alpha-$) 
modifications published in the astronomical literature earlier, showed 
band maxima at different positions (11.4 and 11.8~$\mu$m, respectively). 
These band positions have been taken for characteristics of the 
polytypes and have been used for deriving the polytype of circumstellar 
grains. 

In order to show that this conclusion is fallacious, Fig.\,3 gives 
four other spectra of $\alpha-$ and $\beta-$SiC laboratory samples. 
Clearly, all of these spectra show band maxima longward of 11.8~$\mu$m 
independent of the polytype. Interestingly, the $\beta-$SiC spectra 
peak at the longest wavelengths, around 12.2~$\mu$m. One of the spectra 
(dashed line) reveals a big shoulder towards smaller wavelengths, 
which could indicate a kind of affinity to the $\beta-$SiC spectra 
known from the literature. However, this sample consists of 
pure $\alpha-$SiC, which was proved by X-ray scattering. 

All of these samples consist of submicron grains, especially the 
$\beta-$SiC samples have typical grain sizes of about 50\,nm, well 
below the Rayleigh limit. We will not deal further with size effects 
here, since they have been treated in detail by Andersen et al.\,(1999), 
Mutschke et al.\,(1999) and elsewhere in this volume (Henning \& 
Mutschke). 

The SiC samples have been measured dispersed in KBr. 
This is of course not an ideal preparational method since there 
remains a considerable amount of agglomeration of the grains, which 
together with the grain shape should determine the actual band 
profile. However, we are not aware of a better method for a given 
powder sample. We have to point out that the KBr matrix of course induces an 
effect on the spectrum in the way discussed by Henning \& Mutschke 
(this volume). For SiC there is a possibility to correct for this in 
an approximate way by the method proposed by Papoular et al.\,(1998). 

\section{Measurements on Shape Effects}

\begin{figure}
\plotfiddle{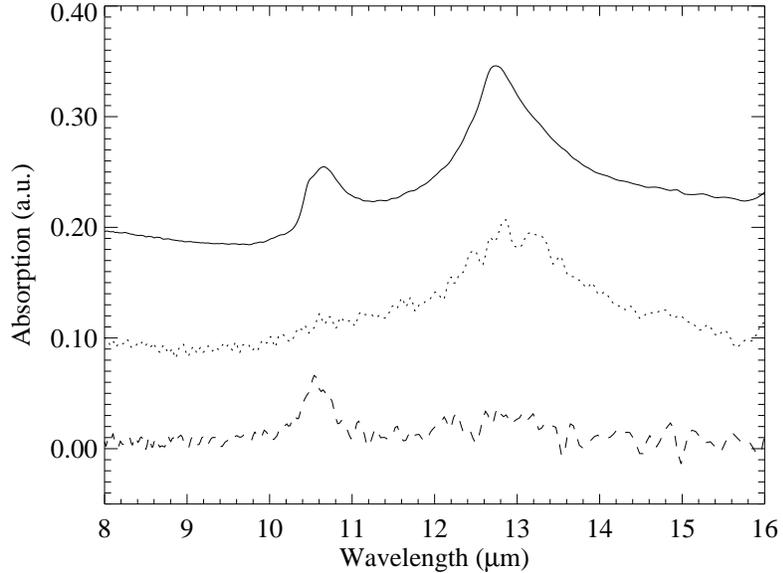}{3in}{0}{70}{70}{-150}{0}
\caption{Infrared microscope spectra of $\beta$-SiC whiskers 
prepared on a silicon window. Solid-line: unpolarized light, 
unaligned whiskers. Dotted line: Radiation polarized parallel 
to whisker axis, dashed line: perpendicular to whisker axis} \label{fig-4}
\end{figure}

As we have seen in the previous section, there is no clear 
dependence of the SiC phonon band profile on the polytype, 
as expected. Nevertheless, the spectra show unsystematic differences in 
peak position and band shape. Following Sect.\,2, a major reason 
for this behaviour could be different grain shapes. 

The problem with discussing grain shape effects arises both from 
the theoretical and experimental side. As pointed out already, there 
is no theory which clearly can predict what band profiles have 
to be expected from a sample of the usually irregular, sharp-edged 
SiC grains. Further, the problem is considerably complicated by the 
fact that in the measurement the particles are not isolated from each 
other but form aggregates around the grains of the embedding medium 
(usually KBr). So far, there is no method to deagglomerate submicron 
powders for an IR measurement. Therefore, we have to expect that 
the probably very elongated aggregates determine the band profile 
to a large extent. This could be the reason why we find usually 
a band maximum quite close to the volume mode. There is no conclusive 
explanation so far for the small shoulder seen in many commercial 
SiC powders at about 10.6~$\mu$m (see the dotted line in Fig.\,3, the 
big bump in the dashed line will be discussed in the next section). 

Fortunately, there is a possibility to demonstrate by a measurement 
at least that shape effects indeed are real. This possibility is 
provided by commercial $\beta$-SiC samples which consist of whiskers 
with a diameter of 1-2~$\mu$m and a length of several ten micrometers. 
According to Sect.\,2, such grains should produce two resonances. One at 
about 10.5~$\mu$m should be caused by electric fields normal to the 
whisker axis, and field components parallel to the whisker axis should 
produce a resonance close to the volume mode. 
Fig.\,4 shows such measurements carried out with an infrared microscope. 
The unpolarized spectrum (solid line) indeed reveals both features 
expected. The IR microscope additionally allowed us to find some 
whiskers which by chance were aligned in the directions of their axes. 
Putting a polarizer directly in front of the sample we have been 
able to measure both resonances separately which proves their origin 
from the special grain shape.

\section{Influence of Impurities}

\begin{figure}
\plotfiddle{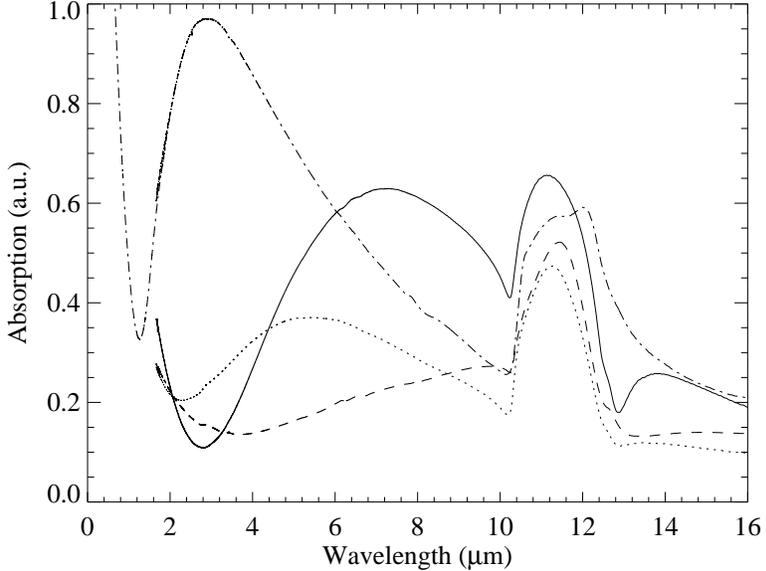}{3in}{0}{70}{70}{-150}{0}
\caption{Spectra of different SiC samples showing very broad surface 
plasmon absorption shortward of the phonon feature. Coupling of plasmons 
and phonons leads to a modification of the phonon band profile with shift 
of the band maximum towards shorter wavelengths. } \label{fig-5}
\end{figure}

The calculations and measurements shown in the previous sections 
should have proven that $\alpha-$ and $\beta-$SiC cannot be 
distinguished in the way adopted by Speck et al.\,(1999). All 
measured spectra shown so far independently of the polytype peaked 
at 11.9-12.2~$\mu$m. It remains to clarify what is the reason for 
some SiC bands in the literature to appear with maxima at about 
11-11.5~$\mu$m. 

For this sake, Fig.\,5 displays four spectra measured on other 
$\alpha$- and $\beta$-SiC samples. Three of them have their phonon band 
maxima in the mentioned wavelength range, the fourth (shown already in 
Fig.\,3) a big shoulder at these short wavelengths. 
Even more striking, all spectra reveal a broad absorption shortward 
of the phonon band reaching down to the near infrared. The maximum 
of the very broad absorption varies between 3 and 10~$\mu$m. These 
features point towards highly damped surface plasmon absorption 
caused by free charge carriers. The resonance frequency of the absorption 
band depends on the plasma frequency which is proportional to the 
square root of the carrier density. Similar absorption spectra are known 
from doped CdO and GaAs small particles. It is also known that 
the plasmon generally couples to the longitudinal phonon and, therefore, 
changes the appearance of the former band. 

To our experience, such a plasmon absorption is quite common to many 
commercial and laboratory SiC products. Unfortunately, the nature of the 
charge carriers is still only partly understood. 
Mutschke et al.\,(1999) showed that plasmon absorption can be introduced 
into SiC particle spectra by doping the SiC grains with nitrogen. These 
investigations will be continued because the induced strong mid-infrared 
absorption should be important for the thermal behaviour of the 
circumstellar SiC particles. If such doped particles would be present 
in circumstellar environment they would probably be quite warm, comparable 
to carbon grains and much warmer than undoped SiC grains. This would also 
give them an advantage for being observed. 

\section{Conclusion}
In our discussion of the optical properties of different polytypes
of SiC we have demonstrated both theoretically and by experimental
studies that it is {\it not} possible to distinguish by IR
spectroscopy between $\alpha$- and $\beta$-SiC dust grains.
Observed profiles of SiC phonon bands contain information mainly
about the morphology (size and shape) of the particles and possibly
about the conductivity of the material. The evaluation of these grain
properties, however, encounters the very general unsolved problem of
calculating band profiles for complicated morphologies. 

Therefore, progress in interpretation of observed band profiles 
in the future will mainly rely on laboratory measurements. The first 
problem which has to be solved in this respect is the preparation 
of particle samples free from agglomeration and with as little 
matrix-influence as possible. The second will be the preparation 
of realistic circumstellar-dust analogues by condensation experiments. 
Both things have been started already for carbonaceous grains by means 
of molecular beam techniques (Schnaiter et al.\,1998) and are our 
program for the next years also with SiC.

\acknowledgments
H.M. gratefully acknowledges support by the Max Planck society. 
This project was partly supported by DFG grant Mu 1164/3-1.

\end{document}